\documentstyle[multicol,aps,pre,epsf]{revtex}
\begin{document}
\title{Corrections To Scaling in Phase-Ordering Kinetics}
\author{A. J. Bray, N. P. Rapapa
and S. J. Cornell}
\address{Department of Physics and Astronomy, The University, 
Manchester M13 9PL, UK}
\date{September 19th, 1997}
\maketitle

\begin{abstract}
The leading correction to scaling associated with departures of the 
initial condition from the scaling morphology is determined for some 
soluble models of phase-ordering kinetics. 
The result for the pair  correlation function has the form 
$C(r,t) = f_0(r/L) + L^{-\omega} f_1(r/L) + \cdots$, 
where $L$ is a characteristic length scale extracted from the energy. 
The correction-to-scaling exponent $\omega$ has the value $\omega=4$ for the 
$d=1$ Glauber model, the $n$-vector model with $n=\infty$, and 
the approximate theory of Ohta, Jasnow and Kawasaki. For the approximate 
Mazenko theory, however, $\omega$ has a non-trivial value: 
$\omega = 3.8836\cdots $ for $d=2$, and $\omega = 3.9030\cdots$ for $d=3$. 
The correction-to-scaling functions $f_1(x)$ are also calculated. 
\end{abstract}
\begin{multicols}{2}

\section{Introduction}
The phenomenon of phase-ordering kinetics has attracted considerable 
attention in recent years \cite{Review}. 
It is now well established that, except in 
certain exceptional circumstances, a scaling regime develops at late 
times, in which the order-parameter morphology is time-independent up 
to one overall time-dependent length scale $L(t)$. This means, for 
example, that the equal-time pair correlation function of the order 
parameter, $C(r,t)=\langle\phi({\bf x}+{\bf r},t)\phi({\bf x},t)\rangle$ 
has the asymptotic scaling form $C(r,t) = f[r/L(t)]$. Relatively little 
attention, however, has been devoted to the study of how the scaling regime 
is approached, i.e.\ the form of `corrections to scaling' in phase-ordering 
kinetics. In particular, an understanding of corrections to scaling is 
important in interpreting experimental or simulation data, and in extracting 
asymptotic scaling exponents and functions. 

This article attempts a first systematic study of corrections to scaling. 
We show that there is a correction-to-scaling exponent $\omega$ 
associated with the deviations of the order parameter morphology from the 
scaling morphology, i.e.\ with the fact that the initial state is not in 
general the scaling state. Presupposing a suitable definition of $L(t)$, 
which will require some discussion, the leading corrections to scaling 
in the pair correlation function enter in the form $C(r,t) = f_0(r/L)
+ L^{-\omega}f_1(r/L)$, where $f_0(x)$ is the `scaling function' and 
$f_1(x)$ the `correction-to-scaling function'. There are indications 
that $\omega$ is, in general, a {\em nontrivial exponent} of phase-ordering 
kinetics. Although for most of the simple, exactly soluble models presented 
here it has the value $\omega=4$, we find that it takes nontrivial values 
within an approximate calculation for more realistic models. 
The models considered 
are the one-dimensional (1D) Glauber model, the $O(n)$ non-linear sigma model 
for $n=\infty$, the 1D time-dependent Ginzburg-Landau (TDGL) equation, 
the Ohta-Jasnow-Kawasaki (OJK) approximation \cite{OJK} 
for the general TDGL equation, 
and lastly the Mazenko approximation \cite{Mazenko}. 
It is the last of these that yields 
non-trivial values for $\omega$. In fact $\omega$ enters the theory in a 
similar way to the exponent $\bar{\lambda}$, which describes the decay of the 
autocorrelation function and is known to be nontrivial \cite{LiuMaz}.
For the 1D TDGL equation, we find that the approach to the scaling state is 
much faster than for the other models: for typical initial conditions the 
corrections to scaling vanish exponentially fast as a function of the 
smallest domain size.

It is important to recognize that corrections to scaling can have more than  
one origin. For a scalar order parameter, for example, in which the 
morphology is specified by the locations of the domain walls, 
the scaling regime requires that the 
domain scale $L(t)$ be much larger than the width $\xi$ of the walls. 
Therefore one expects a correction to scaling associated 
simply with the non-zero width of domain walls, and entering as a power 
of $\xi/L$. This is a distinct contribution from the one considered here, 
which is  associated with departures from the scaling morphology in the 
initial condition. This latter contribution will survive even in the thin  
wall limit described by, for example, the Allen-Cahn theory \cite{Review,AC} 
in which infinitely thin interfaces are driven by their local curvature. 
Throughout this paper we restrict ourselves to this thin-wall limit, 
or the corresponding `hard-spin' limit (or non-linear sigma model) 
for a vector order-parameter. This leads to useful simplifications. 

The paper is organized as follows. Section II deals with the 1D Glauber 
model, and introduces some general concepts. The large-$n$ vector model, 
1D TDGL equation, OJK Theory, and Mazenko theory, are covered in sections 
III to VI respectively. Section VII concludes with a discussion and summary. 

\section{The One-Dimensional Glauber Model}
In the continuum limit, the equation of motion for the pair correlation
function $C(r,t)$ has the form \cite{Bray1D}
\begin{equation}
\partial_t C = \partial_r^2 C,\ \ \ \ \ \ \ C(0,t)=1.
\label{EOM}
\end{equation}
The constraint $C(0,t)=1$ can be built in through a source term at $r=0$, 
i.e.\ we modify the equation of motion to  
\begin{equation}
\partial_t C = \partial_r^2 C + A(t) \delta(r),
\label{EOM1}
\end{equation}
where $A(t)$ is chosen to satisfy the constraint. Let us define $L$ to be 
the average domain size. Then for $r \ll L$, we have 
\begin{equation}
C(r,t) = 1 - 2\rho_w|r| = 1-2|r|/L, 
\label{smallr}
\end{equation} where $\rho_w=1/L$ is the density of domain walls, and we 
have allowed for the possibility of either one or zero walls in an interval 
of length $r$. Equation (\ref{smallr}) is then correct to order $r$. 

Integrating (\ref{EOM1}) across an infinitesimal interval around $r=0$, 
and comparing with (\ref{smallr}), gives $A = 4/L$. Putting this into 
(\ref{EOM1}), and writing $C(r,t)$ in terms of $r$ and $L$ in the form 
\begin{equation}
C(r,t) = f(r/L,L)
\end{equation}
gives the following equation for $f(x,L)$:
\begin{equation}
\dot{L} \frac{\partial f}{\partial L } = \frac{1}{L^2} f'' 
+ \frac{\dot{L}}{L} xf' + \frac{4}{L^2} \delta(x)
\label{scaled}
\end{equation}
where dots and primes indicate derivatives with respect to $t$ and $x$ 
respectively.

In the limit $L \to \infty$, we expect $f(x,L)$ to approach the 
$L$-independent scaling function $f_0(x)$. Balancing powers of $L$ on the
right of (\ref{scaled}) then implies $\dot{L}L = {\rm constant}$, i.e.\ 
$L \propto t^{1/2}$. Including the leading correction to scaling, of relative 
order $L^{-\omega}$, we write
\begin{eqnarray}
\label{L}
\dot{L} & = & a/L + b/L^{1+\omega} + \cdots  \\
f(x,L) & = & f_0(x) + L^{-\omega} f_1(x) + \cdots
\label{f}
\end{eqnarray}
in (\ref{scaled}), and equate terms of leading ($L^{-2}$) and next-to-leading 
($L^{-2-\omega}$) order. This gives
\begin{eqnarray}
\label{f_0}
f_0'' + ax f_0' + 4 \delta(x) & = & 0 \\
f_1'' + ax f_1' + a\omega f_1 + bx f_0' & = & 0.
\label{f_1}
\end{eqnarray}

Integrating (\ref{f_0}) with the boundary conditions $f(0)=1$, 
$f'(0+) = -2$ [from (\ref{smallr})] gives (for $x \ge 0$) 
\begin{eqnarray}
     a & = & 2\pi \\
f_0(x) & = & {\rm erfc}\left( \sqrt{\frac{a}{2}}\,x \right)\ .
\end{eqnarray}
Equation (\ref{f_1}) for $f_1$ becomes
\begin{equation}
f_1'' + ax f_1' + a\omega f_1 = \tilde{b}x\exp(-ax^2/2)\ ,
\end{equation}
where $\tilde{b} = b(2a/\pi)^{1/2}=2b$. Writing $f_1(x) = \exp(-ax^2/2)g(x)$ 
gives
\begin{equation}
g'' - axg' + a(\omega-1)g = \tilde{b} x\ .
\label{g}
\end{equation}

What are the boundary conditions on $g(x)$? Clearly $g(0)=0$, because 
$C(0,t)=1$ is already implemented by $f_0(0)=1$. Similarly, the condition 
$\partial_r C(r,t)|_{r=0}=-2/L$, from (\ref{smallr}), is already 
guaranteed by $f_0'(0)=-2$, so $g'(0)=0$. Furthermore, (\ref{g}) then gives 
$g''(0)=0$, implying that the series expansion for $g(x)$ starts at 
$O(x^3)$. Inserting the series solution $g(x) = \sum_{n=3}^\infty g_n x^n$ 
gives $g_3 = \tilde{b}/6$, and the recurrence relation 
$g_{n+2} = [a(n+1-\omega)/(n+1)(n+2)]g_n$ for the higher-order odd 
coefficients, all even coefficients vanishing. In order that $f_1(x)$ 
decrease faster than a power-law for large-$x$, as required on physical 
grounds for initial conditions with only short-range correlations, the 
series expansion for $g(x)$ must terminate. This gives the condition 
$\omega = n+1 = 4,6,8,\cdots$. We conclude that the leading 
correction-to-scaling exponent for the 1-d Glauber model is
\begin{equation}
\omega=4\ ,
\end{equation}
with corresponding correction-to-scaling function
\begin{equation}
f_1(x) = (\tilde{b}/6)\,x^3\,\exp(-ax^2/2)\ .
\end{equation}

\section{The Large-n Limit of the O(n) model}
It is convenient to work with spins of fixed length, i.e.\ with the 
nonlinear sigma model, to avoid additional (uninteresting) corrections 
to scaling associated with the gradual saturation in the length of the 
spins as the coarsening proceeds. The non-conserved dynamics of the 
$O(n)$ non-linear sigma model is described by the equation \cite{NBM}
\begin{equation}
\partial_t \vec{\phi} = \nabla^2 \vec{\phi} + (\nabla \vec{\phi})^2\ 
\vec{\phi}, 
\label{phi}
\end{equation}
corresponding to the equation $\partial_t \vec{\phi} = -\delta F/\delta 
\vec{\phi}$, with free energy $F=\frac{1}{2}\int d^dx\,(\nabla \vec{\phi})^2$, 
subject to the constraint $[\vec{\phi}({\bf x},t)]^2 = 1$. 

In the limit $n \to \infty$ we can replace $(\nabla \vec{\phi})^2$ by 
its mean in the usual way \cite{NBM}. Let us call this mean $\alpha(t)$. 
In the scaling regime, dimensional analysis gives $\alpha(t) = \lambda/2L^2$, 
where $\lambda$ is a constant. The energy density is just 
$\epsilon(t) = \alpha(t)/2 = \lambda/4L^2$. It is, in fact, convenient to 
{\em define} $L(t)$ through this relation for all times $t$ (in the same 
spirit as the 1d Glauber model). Then multiplying (\ref{phi}) by 
$\phi({\bf x}+{\bf r},t)$ and averaging over initial conditions gives 
the equation
\begin{equation}
\frac{1}{2}\,\frac{\partial C}{\partial t} = \nabla^2 C 
+ \frac{\lambda}{2L^2}\,C 
\label{C}
\end{equation}
for the pair correlation function $C(r,t)$. 

In analogy to our treatment of the 1d Glauber model we write 
$C(r,t) = f(r/L,L)$. Then (\ref{C}) becomes the following equation 
for $f(x,L)$:
\begin{equation}
\frac{\dot{L}}{2}\,\frac{\partial f}{\partial L} 
= \frac{1}{L^2}\,\left(f'' + \frac{d-1}{x}\,f' + \frac{\lambda}{2}\,f\right) 
+ \frac{\dot{L}}{2L}\,xf'\ .
\end{equation}
Inserting the forms (\ref{L}) and (\ref{f}) (taking $a=1/2$ without 
loss of generality), and equating coefficients of the terms in $L^{-2}$ 
and $L^{-(2+\omega)}$ as before gives
\begin{eqnarray}
\label{fn_0}
f_0'' + \left(\frac{d-1}{x} + \frac{x}{4}\right)\,f_0' + 
\frac{\lambda}{2}\,f_0 & = & 0 \\
f_1'' + \left(\frac{d-1}{x} + \frac{x}{4}\right)\,f_1' + 
\left(\frac{\lambda}{2}+\frac{\omega}{4}\right)\,f_1 + 
\frac{b}{2}xf_0' & = & 0\ .
\label{fn_1}
\end{eqnarray}

Consider first $f_0$. The two linearly independent solutions have large-$x$ 
behavior $\sim x^{-2\lambda}$ and $\sim \exp(-x^2/8)$. The power-law 
term must be absent for a physical solution corresponding to an initial 
condition with only short-range correlations. Setting 
$f_0(x) = \exp(-x^2/8)\,g_0(x)$ in (\ref{fn_0}) gives 
$g_0'' + [(d-1)/x - x/4]g_0' + (\lambda/2 - d/4)g_0=0$. The boundary 
conditions are $g_0(0)=1$, $g_0'(0)=0$ (the latter being required by 
the differential equation, to avoid a singularity at $x=0$). 
To retain the gaussian tail in $f_0$, the series solution for $g_0$ must 
terminate. This fixes $\lambda = d/2 + n$ ($n=0,1,2,\cdots$), with a  
corresponding set of polynomial solutions for $g_0$, the first two of 
which are $g_0^{(0)}=1$, and $g_0^{(1)} = 1-x^2/4d$. An explicit solution 
of $C$ for a general initial condition shows that these different scaling 
solutions are selected by the small-$k$ behavior of the structure factor, 
$S({\bf k},t)$ [the Fourier transform of $C({\bf r},t)$], at $t=0$. 
The polynomial solution $g_0^{(n)}$ corresponds to an initial condition 
with $S({\bf k},0) \propto (k^2)^n$ for $k \to 0$. A generic initial 
condition, therefore, selects the $n=0$ solution, i.e.\ $\lambda=d/2$ and 
$f_0(x) = \exp(-x^2/8)$.

Consider now the correction-to-scaling function $f_1$. Putting $f_1(x) = 
\exp(-x^2/8) g_1(x)$ in (\ref{fn_1}), with $\lambda=d/2$ and 
$f_0(x) = \exp(-x^2/8)$, gives 
\begin{equation}
g_1'' + \left(\frac{d-1}{x}-\frac{x}{4}\right)\,g_1' + \frac{\omega}{4}\,g_1 
         = \frac{b}{8}\,x^2\ .
\end{equation}
Again one seeks a series solution for $g_1(x)$. The boundary conditions 
$g_1(0)=0=g_1'(0)$ imply that the series has the form 
$g_1(x) = \sum_{n=2}^\infty g_{n}x^n$. Substituting this into the 
differential equation one readily finds that $g_n=0$ for $n$ odd, $g_2=0$, 
$g_4 = b/[32(d+2)]$, and $g_{n+2} = [(n-\omega)/4(n+2)(n+d)]g_n$ for even 
$n \ge 4$. For a physically sensible large-$x$ behavior, the series must 
terminate, just as in the one-dimensional Glauber model. This requires 
$\omega = 4,6,8,\cdots$. For each such $\omega$, there is a corresponding 
correction-to-scaling function. The generic case, corresponding to 
short-range correlations in the initial condition, is the smallest value 
of $\omega$, i.e.\ $\omega=4$, with $g_1(x) = bx^4/32(d+2)$. The   
correction-to-scaling function is then 
\begin{equation}
f_1(x) = \frac{bx^4}{32(d+2)}\,\exp\left(-\frac{x^2}{8}\right)\ .
\end{equation}

The value of $\omega$ for the $n=\infty$ limit, $\omega=4$, is thus 
identical to that of the $1D$ Glauber model, raising the question of 
whether this could be a general result for models with non-conserved 
dynamics. Indeed, the Ohta-Jasnow-Kawasaki theory discussed in section V
gives the same result. The related approach of Mazenko (section VI), 
however, gives a different, and indeed nontrivial, result. In the following 
section, we discuss another model with $\omega \ne 4$. 

\section{The One-Dimensional TDGL Equation}
Another exactly soluble model (in a suitable limit) is the time-dependent 
Ginzburg-Landau (or TDGL) equation in one dimension. This reads
\begin{equation}
\partial_t\phi = \partial_x^2\phi - V'(\phi),
\label{TDGL}
\end{equation}
where $\phi(x,t)$ is the order-parameter field and $V(\phi)$ is a 
potential function, with a symmetric double-well structure. For convenience, 
we can take the minima of $V$ to be at $\phi=\pm 1$. Eq.\ (\ref{TDGL}) 
represents the simple relaxational dynamics $\partial_t\phi = 
-\delta F/\delta\phi$, with free-energy functional $F[\phi] = \int dx\, 
[(\partial_x\phi)^2/2 + V(\phi)]$. In higher dimensions, this equation 
can be reduced to the standard model of curvature-driven growth 
\cite{AC,Review}, whereby the domain walls move with a velocity proportional 
to their local curvature. In one dimension, of course, the domain walls are 
points, and domain coarsening is driven by the exponentially weak forces 
between adjacent walls, mediated by the exponential tails of the 
domain-wall profiles. 

For the case of interest, where the typical domain size $L$ is large 
compared to the width $\xi$ of a wall, the closest pair of domain walls 
annihilate while the remaining walls hardly move at all, leading to the 
following simplified model \cite{KN}. The smallest domain is combined with 
its two neighbors to form a new single domain. This process is then repeated. 
Eventually the system reaches a scaling state in which the distribution 
$P(l,a)$ of domain sizes $l$, when the smallest domain size is $a$, 
has the scaling form $P(l,a) = a^{-1}f(l/a)$. 

An important feature of this process is that if the domain sizes are 
initially uncorrelated they remain uncorrelated, because the merging 
of three domains into a single domain introduces no correlations 
\cite{BDG,RB}. It is then straightforward to derive an equation of motion 
for the evolution of the distribution $P(l,a)$ as $a$ increases 
\cite{KN,BDG,RB}: 
\begin{eqnarray}
\partial P/\partial a = - P(a,a)\delta(l-a) +\theta(l-3a)\,P(a,a)\times
\phantom{l'}&&\cr
\times\int_a^{l-2a}P(l',a)P(l-l'-a,a)\,dl'&&.
\label{evolution}
\end{eqnarray}
Introducing the Laplace transform with respect to $l$, 
$\phi(p,a)= \int_0^\infty P(l,a)\exp(-pl)\,dl$, gives 
\begin{equation}
\partial \phi/\partial a = -P(a,a)\exp(-pa)\,(1 - \phi^2).
\label{Laplace}
\end{equation}
This equation can, in fact, be integrated exactly but, in the spirit 
of the previous sections, we shall first look for solutions of the form 
\begin{eqnarray}
\label{Laplace1}
\phi(p,a) & = & \phi_0(pa) + \frac{1}{a^\omega}\,\phi_1(pa) + \cdots \\
P(a,a) &=& \frac{1}{a}f_0(1) + \frac{1}{a^{1+\omega}}f_1(1) + \cdots,
\label{Laplace2}
\end{eqnarray}
where $f_0(x)$, $f_1(x)$ are the scaling and correction-to-scaling 
functions for the domain-size distribution, $\phi_0$, $\phi_1$ are the 
corresponding Laplace transforms, and $\omega$ is the correction-to-scaling 
exponent as usual. Inserting (\ref{Laplace1}) and (\ref{Laplace2}) into 
(\ref{Laplace}), and equating leading and subleading terms gives
\begin{eqnarray}
\frac{d\phi_0}{ds} &=& -f_0(1)\frac{\exp(-s)}{s}(1-\phi_0^2), \\
\frac{d\phi_1}{ds} &=& \left\{\frac{\omega}{s} + 2f_0(1)
\frac{\exp(-s)}{s}\phi_0\right\}\phi_1 \cr
&&\phantom{\frac{\omega}{s}}- f_1(1)\frac{\exp(-s)}{s}(1-\phi_0^2),
\end{eqnarray}
where $s=pa$. With the boundary conditions $\phi_0(\infty) = 0$ (which 
follows from the definition of $\phi_0$) and $\phi_1(0)=0$ (which follows 
from the normalization of $P(l,a)$), these equations can be integrated to 
give 
\begin{eqnarray}
\label{scaling}
\phi_0(s) &=& \tanh\left[f_0(1)\int_s^\infty 
\frac{dt}{t}\,\exp(-t)\right], \\
\phi_1(s) &=& f_1(1)s^\omega\left(1-\phi_0^2(s)\right)\int_s^\infty 
\frac{dt}{t^{1+\omega}}\,\exp(-t).
\label{cts}
\end{eqnarray}

Consider first the scaling function $\phi_0(s)$. For small $s$, one 
obtains $\phi_0(s) = 1 - 2e^{2f_0(1)\gamma}\,s^{2f_0(1)} + \cdots$, where 
$\gamma = 0.577\cdots$ is Euler's constant. 
But, from the definition of $\phi_0(s)$ as the Laplace transform of 
$f_0(x)$, one also has $\phi_0(s) = 1 - s\langle x \rangle_0 + \cdots$, 
where $\langle x \rangle_0$ is the mean domain size (in units of the 
minimum domain size $a$) for the scaling distribution. Comparing these 
expansions gives $f_0(1) \le 1/2$, with $f_0(1)=1/2$ when the first moment 
of $f_0(x)$ exists, and $f_0(1)<1/2$ when it doesn't. We shall primarily 
consider the former case, appropriate to an initial distribution with a 
finite first moment. 

Now consider the small-$s$ behavior of the correction-to-scaling function 
$\phi_1(s)$. The factor $(1-\phi_0^2(s))$ in (\ref{cts}) behaves as 
$s^{2f_0(1)}$ for $s \to 0$, i.e.\ as $s$ for $f_0(1)=1/2$. For this case,
the small-$s$ expansion of $\phi_1(s)$ contains a nonanalytic term 
$s^{1+\omega}$ for any $\omega >0$ (with a $\ln s$ factor when $\omega$ 
is an integer). This in turn implies that the correction-to-scaling 
function $f_1(x)$ has the power-law tail $x^{-(2+\omega)}$. This means  
that if the domain-size distribution is given by the scaling 
distribution $f_0(x)$ plus a small perturbation with a power-law tail 
$x^{-(2+\omega)}$, the perturbation is {\em irrelevant} (i.e.\ decays to 
zero) for $\omega > 0$, and {\em the correction-to-scaling exponent is 
$\omega$}. More generally, any initial distribution 
with a power-law tail $x^{-(2+\omega)}$, with $\omega>0$, flows to the 
scaling distribution $f_0(x)$ with $f_0(1) = 1/2$. 
For $\omega < 0$, the perturbation is relevant. 
In this connection one should note that the scaling 
distributions with $f_0(1) < 1/2$ have the power-law tail $x^{-[1+2f_0(1)]}$, 
i.e.\ they decay more {\em slowly} than $1/x^2$. So any initial condition 
that falls off more quickly than one of these `special' fixed distributions 
flows to the `generic' fixed distribution with $f_0(1)=1/2$ (which may  
easily be shown to have an exponential tail).

We have shown that any finite correction-to-scaling exponent $\omega$ 
corresponds to a correction-to-scaling function with a power-law tail. 
This implies that corrections to scaling that decay faster than any  
power-law in the scaling variable $x$  have an infinite correction 
to scaling exponent, i.e.\ that the amplitude of the correction does 
not have the power-law form $a^{-\omega}$ assumed in (\ref{Laplace1}). 
To clarify this case, we construct the exact solution of (\ref{Laplace}) 
for an arbitrary initial condition: 
\begin{equation}
\phi(p,a) = \tanh\left[\int_a^\infty da'\,P(a',a')\,\exp(-pa')\right].
\label{exact}
\end{equation}
The arbitrary function of $p$ that formally appears inside the argument 
of the $\tanh$, as an additive integration constant,  must vanish because 
$\phi(p,\infty) = 0$. The function $P(a,a)$ appearing in (\ref{exact}) 
is determined by the initial condition, for which $a=a_0$, through the 
inverse Laplace transform
\begin{equation}
P(a,a) = \int_C \frac{dp}{4\pi i}\,\exp(pa)\,\ln\left(\frac{1+\phi(p,a_0)}
{1-\phi(p,a_0)}\right),
\label{Bromwich}
\end{equation}
where the contour $C$ runs parallel to the imaginary axis, to the right of 
any singularities of the integrand. 

Equations (\ref{exact}) and (\ref{Bromwich}) represent an exact closed-form 
solution for the full evolution from an arbitrary initial condition. 
For explicit results, however, one needs to be able to evaluate the Bromwich 
integral (\ref{Bromwich}). For illustrative purposes, we consider two 
simple initial conditions. The first is the simple exponential 
distribution $P(l,0) = \mu\exp(-\mu l)$, with Laplace transform 
$\phi(p,0) = \mu/(p+\mu)$ (i.e.\ $a_0=0$ here). Inserting this in 
(\ref{Bromwich}) gives 
\begin{equation}
P(a,a) = \frac{1}{2a}[1-\exp(-2\mu a)].
\label{P}
\end{equation}
One sees that $P(a,a)$ approaches the scaling limit $1/2a$ exponentially 
fast as a function of $a$, i.e.\ faster than any power. This is in accord 
with our result that a power-law correction to scaling is associated with 
power-law tails in the initial domain-size distribution. Inserting (\ref{P}) 
into (\ref{exact}), and expanding for $\mu a \gg 1$, gives 
\begin{equation}
\phi(p,a) = \phi_0(s) - \frac{e^{-2\mu a-s}}{4\mu a}
\left(1-\phi_0(s)^2\right) + \cdots,
\end{equation}
where $\phi_0(s)$ (with $s=pa$) is the scaling function. Again, the 
correction is exponentially small. 

Our second example shows that the correction to scaling can be an 
oscillatory function of $a$. For $P(l,0)=\mu^2l\exp(-\mu l)$, 
one obtains $\phi(p,0) = \mu^2/(p + \mu)^2$. Inserting this into 
(\ref{Bromwich}) gives $P(a,a) = [1 - 2\cos(\mu a)\exp(-\mu a)
+ \exp(-2\mu a)]/2a$, so that the leading correction to scaling is 
the term involving $\cos(\mu a)$. The difference between this case 
and the first example is that in the former the singularities of the function 
$h(p) \equiv \ln\{[1+\phi(p,0)]/[1-\phi(p,0)]\}$ appearing in (\ref{Bromwich}) 
are branch points on the real axis at $p=0$ and $-2\mu$, whereas in the 
latter they are branch points at $p=0$, $-(1 \pm i)\mu$ and $-2\mu$. 
The leading scaling part $1/2a$ in $P(a,a)$ is associated with the 
singularity of $h(p)$ at $p=0$. The leading correction to scaling derives 
from the singularity with the largest real part (excluding $p=0$). If this 
singularity is off the real axis, the leading correction to scaling will be 
oscillatory. In all the other models considered in this paper, however, 
the leading correction to scaling is monotonically decreasing function of 
time. 

\section{The OJK Theory}
\label{section:OJK}
The theory of Ohta, Jasnow, and Kawasaki (OJK) \cite{OJK} for the pair 
correlation function of a nonconserved scalar field starts from the 
Allen-Cahn equation relating the interface velocity to its local 
curvature \cite{AC}. The theory is expressed in terms of a `smooth' 
auxiliary field $m({\bf x},t)$ whose zeros give the positions of the 
interfaces between domains of the two phases. The field $m$ obeys 
the diffusion equation $\partial m/\partial t = \nabla^2 m$ (we absorb the 
diffusion constant into the timescale). The initial conditions are taken to 
be  is taken to be gaussian. The normalized pair correlation function of 
$m$ is
\begin{equation}
\gamma(r,t) = \frac{\langle m({\bf x},t)\,m({\bf x}+{\bf r},t)\rangle}
{\langle m({\bf x},t)^2 \rangle}\ ,
\end{equation}
while the correlation function for the order-parameter field $\phi$ is 
\begin{equation}
C(r,t) = \langle {\rm sgn}[m({\bf x},t)]\,{\rm sgn}[m({\bf x}+{\bf r},t)]
\rangle  = \frac{2}{\pi}\sin^{-1}\gamma(r,t)\ .
\label{OJK}
\end{equation}     
Note that by using the sgn function to relate the order parameter to the 
auxiliary field we are working in the `thin wall' limit. This means that 
we are neglecting any corrections to scaling associated with the finite 
width of the domain walls, and are focusing instead on corrections deriving  
from the initial conditions. This is the same approach that we are using  
throughout this paper.

It is simplest, in the first instance, to compute the corrections to scaling 
associated with the function $\gamma(r,t)$. If we define 
$h(r,t) = \langle m({\bf x},t)\,m({\bf x}+{\bf r},t)\rangle$, then 
$\gamma(r,t) = h(r,t)/h(0,t)$. From the diffusion equation for $m$, one 
obtains immediately $(1/2)\partial_t h = \nabla^2 h$, giving 
$(1/2)\partial_t \gamma = \nabla^2 \gamma + \alpha(t)\gamma$, where 
$\alpha(t) = - [\partial_t \ln h(0,t)]/2$. 
Since $h$ satisfies a diffusion equation, we know 
that for a short-ranged initial condition, peaked around $r=0$, 
$h(0,t) \sim t^{-d/2}$ asymptotically, giving 
$\alpha(t) \to d/4t$ for large $t$. 
It is convenient, however, to incorporate corrections to scaling in $L(t)$ 
through the requirement that $\alpha(t) = \lambda/2L^2$ exactly, i.e.\ 
$(1/2)\partial_t \gamma = \nabla^2 \gamma + (\lambda/2L^2)\gamma$. 
Choosing the scale of $L$ such that $L \to t^{1/2}$ asymptotically then 
fixes $\lambda=d/2$, as in the large-$n$ theory. Writing 
$\gamma(r,t) = \gamma_0(r/L) + L^{-\omega}\gamma_1(r/L)$, and 
$\dot{L} = 1/2L + b/L^{1+\omega}$, as usual, substituting into the 
equation for $\gamma$, and equating leading and next-to-leading powers of 
$L$, gives 
\begin{eqnarray}
\label{gamman_0}
\gamma_0'' + \left(\frac{d-1}{2} + \frac{x}{4}\right)\,\gamma_0' + 
\frac{d}{4}\,\gamma_0 & = & 0 \\
\gamma_1'' + \left(\frac{d-1}{2} + \frac{x}{4}\right)\,\gamma_1' + 
\left(\frac{d+\omega}{4}\right)\,\gamma_1 + 
\frac{b}{2}x\gamma_0' & = & 0\ .
\label{gamman_1}
\end{eqnarray}
In fact these equations are identical to equations (\ref{fn_0}) and 
(\ref{fn_1}) of the large-$n$ theory, with the same boundary conditions, 
and the results for $\gamma_0$ and $\gamma_1$ can be simply read off:
\begin{eqnarray}
\label{g0}
\gamma_0(x) = \exp(-x^2/8), \\
\gamma_1(x) = [bx^4/32(d+2)]\,\exp(-x^2/8).
\label{g1}
\end{eqnarray}
The correction-to-scaling exponent is $\omega=4$, as in the large-$n$ theory. 

The order-parameter correlation function $C(r,t)$ has the expansion 
$C(r,t) = f_0(r/L) + L^{-\omega}f_1(r/L) + \cdots$. The functions $f_0(x)$ 
and $f_1(x)$ follow immediately from (\ref{OJK}), (\ref{g0}) and (\ref{g1}):
\begin{eqnarray}
f_0(x) & = & \frac{2}{\pi}\,\sin^{-1}\gamma_0(x), \\
f_1(x) & = & \frac{2}{\pi}\,\frac{\gamma_1(x)}{[1-\gamma_0^2(x)]^{1/2}}.
\end{eqnarray}
Note that $f_1(x) \sim x^3$ for small $x$, so that the linear term in $x$ 
from $f_0$ is not modified -- our definition of $L$ has ensured that
areal density of domain walls, and therefore the energy, varies as $1/L$ 
exactly.  

\section{The Mazenko Theory}
\label{section:Mazenko}
An alternative `gaussian closure' theory to that of OJK has been proposed 
by Mazenko \cite{Mazenko}. For nonconserved scalar fields, the pair 
correlation function $C(r,t)$ satisfies the closed equation 
\begin{equation}
\frac{1}{2}\frac{\partial C}{\partial t} = \nabla^2 C + \frac{1}{\pi S_0(t)} 
\,\tan\left(\frac{\pi}{2}\,C \right)\ .
\label{Mazenko}
\end{equation}
The function $S_0(t)$ is defined as $\langle m({\bf r},t)^2 \rangle$, 
where $m$ is a gaussian auxiliary field as before. For present purposes, 
however, it is sufficient to note that $S_0$ has dimensions 
$({\rm length})^2$. In fact, it is convenient to {\rm define} the 
coarsening length scale $L(t)$ by $S_0 = L^2/\lambda$, where $\lambda$ 
is a constant whose value is fixed by physical requirements \cite{Mazenko}. 
This definition of $L$ is in accord with our previous definitions, as we 
shall see.

\begin{figure}[t]
\narrowtext
\epsfxsize=\hsize
\epsfbox{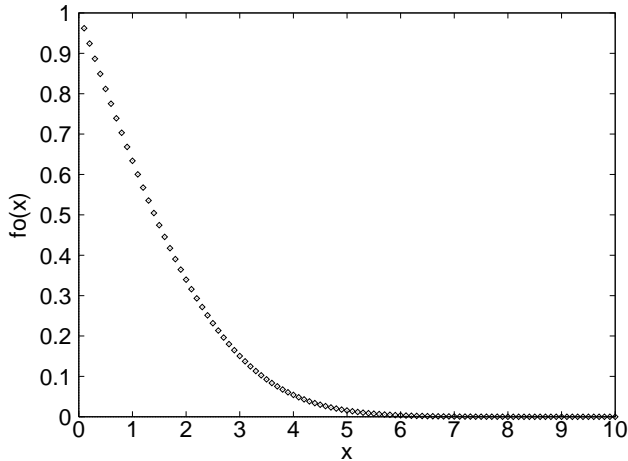}
\caption{The scaling function $f_0(x)$ in the Mazenko theory for $d=2$.}
\end{figure}

Writing $S_0=L^2/\lambda$ in (\ref{Mazenko}), setting $C(r,t) = f_0(r/L) 
+ L^{-\omega}f_1(r/L) + \cdots$,  $dL/dt = 1/2L + b/L^{1+\omega} + \cdots$, 
and equating leading and next-to-leading powers of $L$ in the usual way gives 
the following equations for the functions $f_0(x)$ and $f_1(x)$:
\begin{eqnarray}
\label{M1}
f_0'' + \left(\frac{d-1}{x}+\frac{x}{4}\right)f_0' + 
\frac{\lambda}{\pi}\,\tan\left(\frac{\pi}{2}\,f_0\right) & = & 0 \\
f_1'' + \left(\frac{d-1}{x}+\frac{x}{4}\right)f_1' + 
\frac{\lambda}{2}\,\sec^2\left(\frac{\pi}{2}\,f_0\right)f_1 +&& \cr
+\frac{\omega}{4}\,f_1 + \frac{b}{2}xf_0'& = & 0\ .
\label{M2}
\end{eqnarray}
Eq.\ (\ref{M1}) is to be solved with boundary conditions $f_0(0)=1$, 
$f'(0) = -(1/\pi)[2\lambda/(d-1)]^{1/2}$, the latter following from
a small-$x$ analysis of the equation. The parameter $\lambda$ is 
determined as follows \cite{Mazenko}. For large $x$, (\ref{M1}) reduces 
to the linear equation $f_0'' + [(d-1)/x + x/4]f_0' + (\lambda/2)f_0=0$. 
This equation has the same form as the large-$n$ equation (\ref{fn_0}), 
and the two linearly independent large-$x$ solutions have the large-$x$ 
forms $x^{-2\lambda}$ and $\sim \exp(-x^2/8)$ discussed in section III. 
If the equation is integrated forward from $x=0$, for general $\lambda$ 
a linear combination of the two large-$x$ solutions will be obtained. 
But a power-law decay is unphysical if the initial conditions contain 
only short-range spatial correlations. The parameter $\lambda$ is 
fixed by the requirement that this unphysical power-law decay be absent. 
Imposing this condition through a numerical solution of (\ref{M1}) gives 
$\lambda = 0.711277$ for $d=2$, and $\lambda = 1.327411$ for 
$d=3$ (correct to the number of figures given). The exponent $\bar{\lambda}$, 
which describes the decay of the autocorrelation function via 
$\langle \phi({\bf x},t)\phi({\bf x},0)\rangle \sim L(t)^{-\bar{\lambda}}$, 
is given by $\bar{\lambda}=d-\lambda$ within this theory \cite{LiuMaz}. 
For $d \to \infty$, the OJK solution is recovered and $\lambda \to d/2$.

\begin{figure}[t]
\narrowtext
\epsfxsize=\hsize
\epsfbox{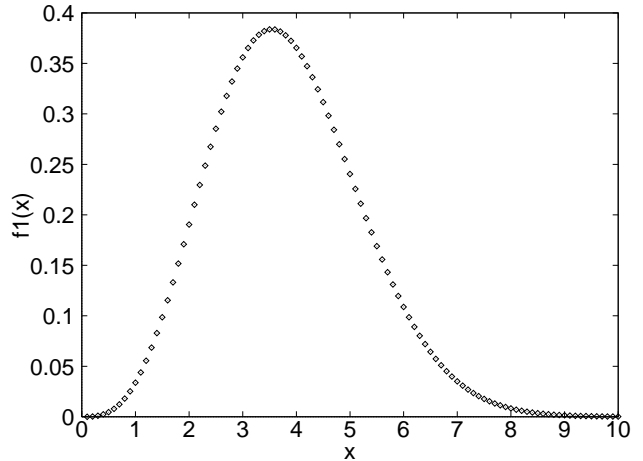}
\caption{The correction-to-scaling function $f_1(x)$ in the Mazenko theory for
$d=2$.}
\end{figure}

The correction-to-scaling $\omega$ is determined in a similar way from 
Eq.\ (\ref{M2}). The boundary conditions on this equation are 
$f_1(0) = 0 = f_1'(0)$. For large-$x$, the homogeneous part of this equation 
(i.e.\ without the final term) reduces to the homogeneous part of the 
equivalent large-$n$ equation (\ref{fn_1}). The two linearly independent 
large-$x$ solutions decay as $\sim \exp(-x^2/8)$ and 
$\sim x^{-2\lambda - \omega}$. The exponent omega is determined by the 
condition that the power-law decaying term is absent from the large-$x$ 
solution. Implementing this condition numerically gives 
$\omega = 3.8836$ for $d=2$ and $\omega = 3.9030$ for $d=3$. 
For $d \to \infty$, the OJK result $\omega = 4$ is recovered.

Figures 1 and 2 show the scaling and correction-to-scaling functions 
for $d=2$ (those for $d=3$ are very similar). The {\em amplitude} of the 
correction to scaling function is fixed by the constant $b$ in equation 
(\ref{M2}). The value $b=2$ was used in Figure 2. 

Note that, within the Mazenko approach, the exponent $\omega$ is 
{\em nontrivial}. Indeed in this theory $\omega$ is on the same footing 
as $\bar{\lambda}$, which is known to be nontrivial in general. This suggests 
that, like $\bar{\lambda}$, the correction-to-scaling exponent $\omega$ is 
in general a nontrivial exponent of phase-ordering kinetics.

\section{Conclusion}
An understanding of the corrections to scaling is important in analysing 
data from experiments or simulations. In this paper we have computed the 
correction-to-scaling exponents $\omega$, and the corresponding 
correction-to-scaling functions, for a number of models of phase-ordering 
dynamics. For simplicity we have focussed on the correction 
to scaling associated with the approach of the order-parameter morphology 
to its scaling form, i.e.\  the correction arising from a nonscaling 
initial condition, and have suppressed contributions associated with, 
for example, the finite thickness of the domain walls by working in the 
thin-wall limit. 

In simple soluble models such as the ID  Glauber model, the $n=\infty$ model, 
and the OJK theory, the exponent $\omega$ takes an integer value, 
$\omega=4$. Within the Mazenko theory, however, $\omega$ takes nontrivial 
values, suggesting that this exponent is, in general, a {\em new nontrivial 
exponent} of phase-ordering kinetics. The 1D TDGL equation is anomalous in 
that corrections to scaling vanish exponentially fast as a function of the 
smallest domain size. 

It is important to note that the values of $\omega$ computed in this paper 
correspond to a special, though natural, definition of the domain scale $L$ 
through the mean energy density. Specifically the energy density is defined 
to be exactly $1/L$ (up to an overall constant) for the scalar models, i.e.\ 
the areal density of  domain walls is proportional to $1/L$, except for the 
1D TDGL equation where the scale length was conveniently chosen to be the 
size of the smallest domain. For the large-$n$ non-linear sigma model, the 
energy density is defined to be exactly $1/L^2$, again up to an overall 
constant. These definitions of $L$ are optimal in that they minimize the 
corrections to scaling, expressed in terms of $L$, i.e.\ they maximize 
$\omega$. 

In the present paper we have considered systems with nonconserved order 
parameter only. In future work we hope to discuss corrections to scaling 
in systems with conserved order parameter.

\end{multicols}
\end{document}